\pdfoutput=1
\documentclass{aa}  
\usepackage{float}
\usepackage{amsmath}
\usepackage{graphicx}
\usepackage{nicefrac}
\usepackage{txfonts}
\usepackage{xcolor}
\usepackage{url}
\usepackage[colorlinks,linkcolor=blue,citecolor=blue,urlcolor=blue]{hyperref}

\begin{document}

   \title{The Detection of Transiting Exoplanets by \textit{Gaia}}

   \subtitle{}

   \author{Aviad Panahi \inst{1} \and 
   Shay Zucker\inst{2}  \and 
   Gisella Clementini\inst{3} \and
   Marc Audard\inst{4} \and
   Avraham Binnenfeld\inst{2} \and 
   Felice Cusano\inst{3} \and
   Dafydd Wyn Evans\inst{5} \and
   Roy Gomel\inst{1} \and 
   Berry Holl\inst{4} \and
   Ilya Ilyin\inst{6} \and
   Gr\'{e}gory Jevardat de Fombelle\inst{4} \and
   Tsevi Mazeh\inst{1} \and
   Nami Mowlavi\inst{4} \and
   Krzysztof Nienartowicz\inst{4,7} \and
   Lorenzo Rimoldini\inst{4} \and
   Sahar Shahaf\inst{8} \and
   Laurent Eyer\inst{4} 
     }

   \institute{School of Physics and Astronomy, Raymond and Beverly Sackler Faculty of Exact Sciences, Tel Aviv University, Tel Aviv 6997801, Israel
    \and
    Porter School of the Environment and Earth Sciences, Raymond and Beverly Sackler Faculty of Exact Sciences, Tel Aviv University, Tel Aviv 6997801, Israel
    \and
    INAF - Osservatorio di Astrofisica e Scienza dello Spazio di
    Bologna, via Piero Gobetti 93/3, 40129 Bologna, Italy
    \and 
    Department of Astronomy, University of Geneva, Chemin Pegasi 51,
    1290 Versoix, Switzerland
    \and
    Institute of Astronomy, University of Cambridge, Madingley Road,
    Cambridge CB3 0HA, UK
    \and
    Leibniz-Institut für Astrophysik Potsdam (AIP), An der Sternwarte 16, 14482 Potsdam, Germany
    \and
    Sednai Sàrl, Geneva, Switzerland
    \and
    Department of Particle Physics and Astrophysics, Weizmann Institute of Science, Rehovot 7610001, Israel
    }

   \date{Received March 8, 2022; accepted May 19, 2022}

   
   \abstract
{The space telescope Gaia is dedicated mainly to performing high-precision astrometry, but also spectroscopy and epoch photometry which can be used to study various types of photometric variability. One such variability type is exoplanetary transits. The photometric data accumulated so far have finally matured enough to allow the detection of some exoplanets.}
{In order to fully exploit the scientific potential of \textit{Gaia}, we search its photometric data for the signatures of exoplanetary transits.}
{The search relies on a version of the Box-Least-Square (BLS) method, applied to a set of stars prioritized by machine-learning classification methods. An independent photometric validation was obtained using the public full-frame images of \textit{TESS}. In order to validate the first two candidates, radial-velocity follow-up observations were performed using the spectrograph PEPSI of the Large Binocular Telescope (LBT).}
{The radial-velocity measurements confirm that two of the candidates are indeed hot Jupiters. Thus, they are the first exoplanets detected by \textit{Gaia} -- Gaia-1b and Gaia-2b.}
{Gaia-1b and Gaia-2b demonstrate that the approach presented in this paper is indeed effective. This approach will be used to assemble a set of additional exoplanet candidates, to be released in \textit{Gaia} third data release, ensuring better fulfillment of the exoplanet detection potential of \textit{Gaia}.}

   \keywords{Methods: data analysis --  planets and satellites: detection -- techniques: photometric -- techniques: radial velocities}

   \maketitle
%

\section{Introduction}
Transit photometry is currently the most prolific method for detecting exoplanets, with more than $3\,000$ discovered to this day, mostly using space-based missions, like \textit{Kepler} \citep{Borucki2010Kepler} and \textit{TESS} \citep{Ricker2015}. These missions excel in detecting exoplanets thanks to their high cadence, highly-precise photometry and continuous sampling of large samples of stars. Nevertheless, there is still some chance that sparse low-cadence photometry, while far from being optimal for that purpose, would also be able to detect transiting exoplanets.
In fact, the transits of two exoplanets that had been detected by radial velocities -- HD\,209458b \citep{2000ApJ...529L..45C} and HD\,189733b \citep{HD189733b} were later found in the archived photometry of the first all-sky astrometric mission \textit{Hipparcos} \citep{HIPPARCOS}.
The \textit{Hipparcos} photometric time series had fewer than $200$ measurements each, but still managed to sample the planetary transits \citep{HD209458HIPPARCOS, HD189733HIPPARCOS}.
These detections proved it was possible that such sparse and low-cadence observations may sample a meaningful number of transit events.

The current astrometric mission \textit{Gaia} \citep{Gaia2016} has already revolutionized astronomy with its high-precision astrometry for about $1.8$ billion stars. On the other hand, similarly to \textit{Hipparcos}, the photometry produced by \textit{Gaia} is very sparse, with an irregular sampling scheme, which as mentioned above is suboptimal for detecting exoplanetary transits. Still, early on in its first two years of operation \textit{Gaia} did manage to capture some transits of previously-known exoplanets, such as WASP$\,19\,$b \citep{Hebb_2009} and WASP$\,98\,$b \citep{10.1093/mnras/stu410}\footnote{\url{https://www.cosmos.esa.int/web/gaia/iow_20170209}}. \citet{Dzigan_2012} estimated that with five years of \textit{Gaia} photometry (which is more precise than that of \textit{Hipparcos}) it should be possible to detect several hundreds of transiting Jovian exoplanets, as well as brown dwarfs \citep{2021arXiv210902647H}.

As members of the Data Processing and Analysis Consortium (DPAC) of \textit{Gaia}, we hereby present the approach taken by DPAC in order to exploit the potential of \textit{Gaia} to detect transiting exoplanets. We have found $41$ candidates and validated by radial-velocity (RV) follow-up observations the first two candidate planet host-stars: Gaia~EDR3~3026325426682637824 and Gaia~EDR3~1107980654748582144, which we will henceforth refer to as Gaia-1 and Gaia-2, respectively.

Sect.~\ref{sec:methods} describes the search procedure, including validation by \textit{TESS} photometry. Sect.~\ref{sec:confirmed} presents a detailed analysis of the two candidates that we validated with RV measurements. Finally we conclude in Sect.~\ref{sec:conc} and put the results in the context of \textit{Gaia} data releases.

\section{Methods} \label{sec:methods}
Transit-finding algorithms, such as the box-fitting least squares (BLS) algorithm \citep{Kovacs2002}, have a typical time complexity of $\mathcal{O}(N \cdot N_\mathrm{p})$, where $N$ is the number of points in the light curve and $N_\mathrm{p}$ is the number of trial periods scanned.
\textit{Gaia} Early Data Release~3 (EDR3) is the result of analyzing the data of the first $34$ months of \textit{Gaia} operation, and scanning for periods in a range of $[0.5,100]$ days would require about $\mathcal{O}(10^4)$ trial periods. With $1.8$ billion stars in the \textit{Gaia} database, and having a few dozen measurements for each star, applying the conventional search algorithms to all of them would be prohibitively time-consuming and impractical. 
Therefore we decided not to perform an exhaustive transit search on all the observed stars,
but focus on stars that passed an initial examination, as part of a general classification step that used machine learning methods to classify \textit{Gaia} time series into variability classes (Sect.~\ref{sec:classification}). 

\subsection{Gaia photometry}
The light curves we scanned included the combined epoch photometry from all three photometric bands of \textit{Gaia} -- $G$, $G_\mathrm{BP}$ and $G_\mathrm{RP}$, after independently subtracting their median magnitudes. We combined the three bands in an effort to increase the number of samples in each light curve, assuming the transit effect is achromatic (to a first approximation) and would therefore be similar in all three bands. We searched for outliers based on their distance from the median magnitude, in terms of the standard deviation, $\sigma$, and excluded samples that were $2 \sigma$ brighter or $5 \sigma$ fainter than the median.

\subsection{Training set} \label{sec:training}
We compiled a training set consisting of \textit{Gaia} light curves with noticeable transits of previously known exoplanets. We applied a dedicated version we have developed of the BLS algorithm 
to the \textit{Gaia} light curves of all known transiting exoplanets in order to find these transits. This version is scanning a restricted range in the parameter space of the three temporal transit parameters: period, mid-transit time and duration. We use a very similar approach in another study in which we use \textit{Gaia} photometry to test for false positives in the \textit{TESS} detections due to blends with background binaries (Panahi et al., in prep). The scan only covered a range of $\pm 3\sigma$ for each transit parameter, as published in the NASA exoplanet archive \citep{NASA2013}. At the end of each run, a set of preliminary transit parameters was obtained, along with a statistic we dubbed Transit SNR ($\mathcal{SNR}_\mathrm{T}$):
\begin{equation}
    \mathcal{SNR}_\mathrm{T} \equiv \dfrac{d}{\sigma^{}_\mathrm{OOT}} \sqrt{N^{}_\mathrm{IT}} \ ,
\end{equation}
where $d$ is the transit depth found by BLS, $\sigma^{}_\mathrm{OOT}$ is the standard deviation of the out-of-transit (OOT) measurements (as a proxy to the random variability of the whole light curve), and $N_\mathrm{IT}$ is the number of in-transit (IT) points.
We visually inspected the folded light curves of the stars that had $\mathcal{SNR}_\mathrm{T}>6$ and selected the ones with a clear transit-like signal, resulting in $77$ sources to be used as the training set. 

\subsection{Classification} \label{sec:classification}
A general supervised classification module was applied to all variability types (Rimoldini et al., in prep.) that included a generic computationally efficient period search method \citep[Generalised Lomb-Scargle, GLS;][]{2009A&A...496..577Z}, although it was not necessarily optimal for all classes. Given the weak signal of exoplanetary transits and the likely unreliable period GLS obtained for them, the classifier was designed to attempt initial identification of this class from simple epoch photometry statistics in the three \textit{Gaia} bands, without the important test of periodicity.

Consequently, the initial set of $77$ training sources was further trimmed to enhance the clarity of the signal and thus the chances of detection. For example, sources with negative $G$-band standardized skewness (in magnitude) were excluded, as noise fluctuations on the bright side of the time series were larger than the transit signal in those cases. Eventually only $66$ sources were used to train classifiers for exoplanetary transits, among a training set of almost $60\,000$ objects and $40$ classes (before the selection of publishable classes).

Random Forest \citep{Breiman.Random.Forest} and XGBoost \citep{XGBoost_arxiv} classification methods were used to model the training set with a list of attributes defined in section~10.3.3 of \citet{DR3-documentation_short}\footnote{To be made public with \textit{Gaia} Data Release 3 (DR3).}. The XGBoost classifier was distinctively more effective than Random Forest in naturally identifying these rare training objects and it was thus adopted for predicting exoplanetary transits. This resulted in a total number of $18\,383$ candidates.

\begin{figure}
	\includegraphics[width=1\linewidth]{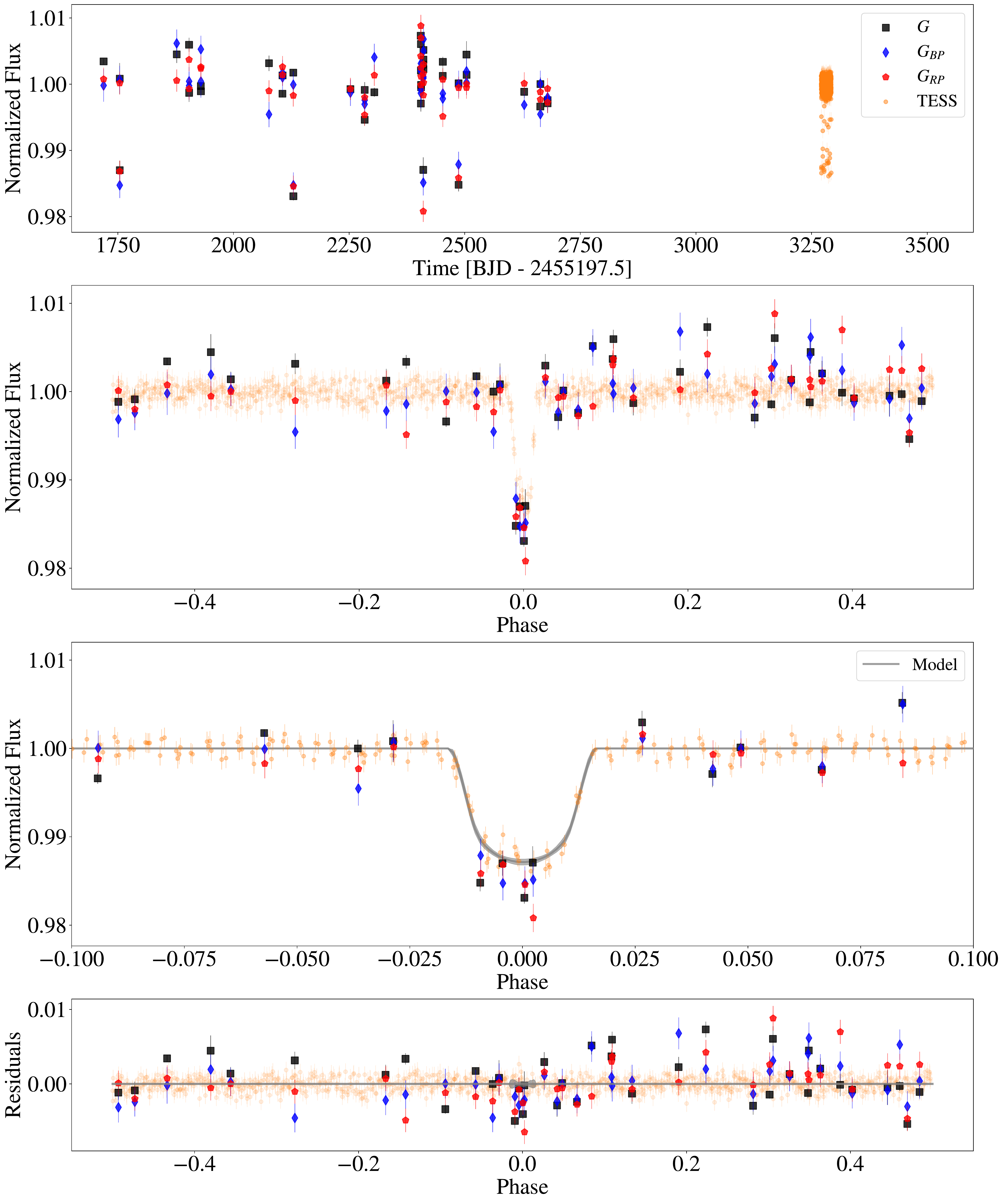}
	\caption{Gaia-1: (\textit{Top}) combined photometry of \textit{Gaia} and \textit{TESS} FFI. (\textit{Second panel}) Phase-folded light curves according to the transit parameters listed in Table~\ref{table:posteriors}. (\textit{Third panel}) Zoom in on transit. (\textit{Bottom}) Residuals showing a possible systematic effect. }
	\label{fig:30263_Phot} 
\end{figure}
\begin{figure}
	\includegraphics[width=1\linewidth]{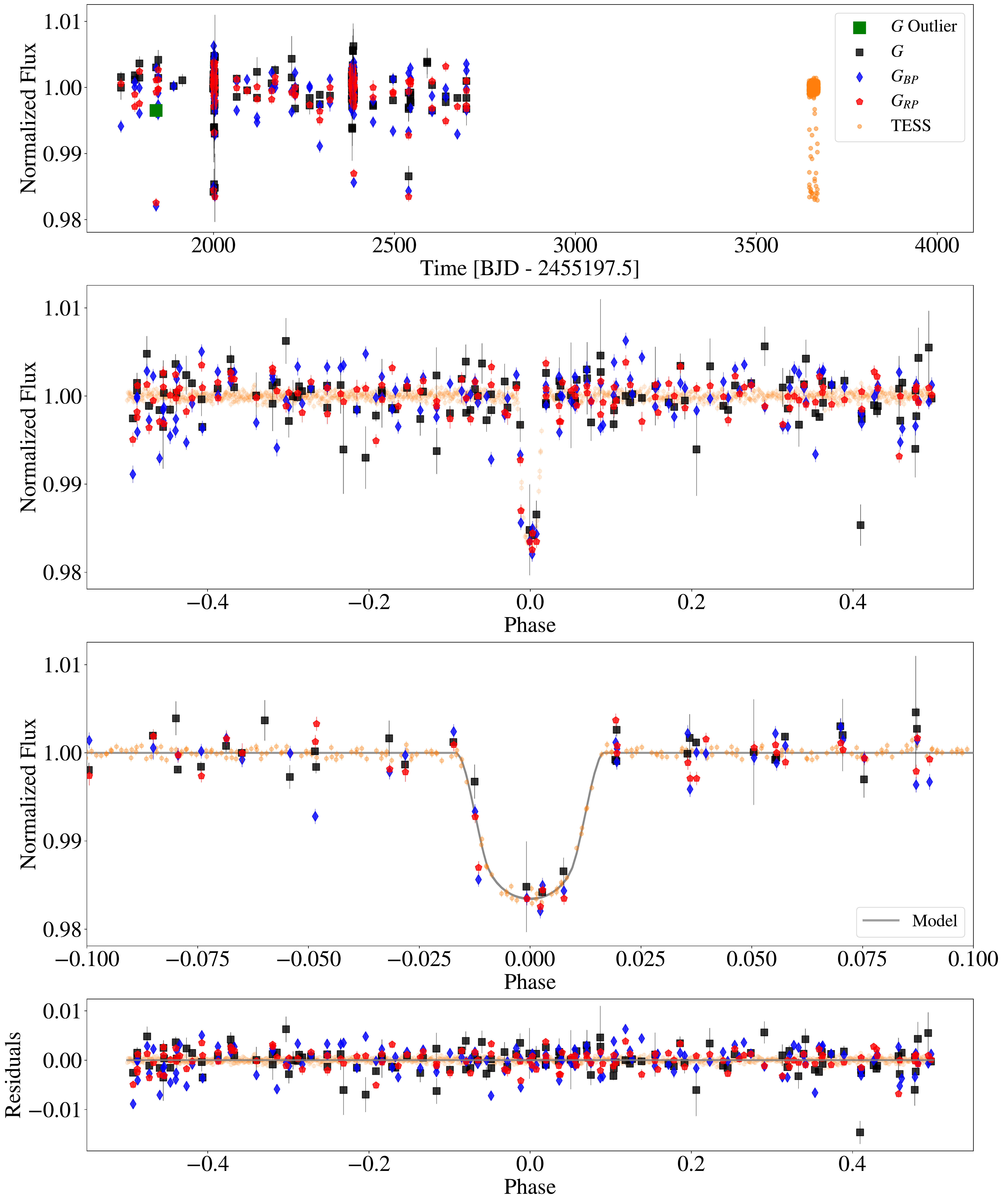}
	\caption{Gaia-2: (\textit{Top}) combined photometry of \textit{Gaia} and \textit{TESS} FFI. (\textit{Second panel}) Phase-folded light curves according to the transit parameters listed in  Table~\ref{table:posteriors}. (\textit{Third panel}) Zoom in on transit. (\textit{Bottom}) Residuals with no visible systematic effects.}
	\label{fig:11079_Phot} 
\end{figure}

\subsection{Initial candidates} \label{sec:init_cand}
We applied to the $18\,383$ initial candidates a dedicated implementation of the BLS algorithm -- SparseBLS \citep{PanahiZucker2021}, which we had developed especially for \textit{Gaia} photometry. Unlike the BLS version that we used to compile the training set (Sect.~\ref{sec:training}), SparseBLS scans only an array of trial periods and estimates the mid-transit time and transit duration from the actual data timestamps, as opposed to a pre-determined grid. SparseBLS is especially suitable for sparse light curves containing hundreds of measurements or fewer, since the run-time depends quadratically on the number of samples in the light curve. 

We scanned with SparseBLS periods in the range of $[0.5,100]$ days, with a frequency step of $\Delta f=10^{-5}\, \textrm{d}^{-1}$.
This resulted in preliminary transit parameters, including period, time of mid-transit and transit depth, along with other BLS statistics, such as the Signal Detection Efficiency (SDE) \citep{Kovacs2002, Alcock2000}, which quantifies the prominence of the periodogram peak. Similarly to BLS, SparseBLS uses the Signal Residue (SR) score for its periodogram. The SR is the part of the sum of squared residuals in a least-square fit that depends on the attempted model. For this score the SDE is simply defined as:
\begin{equation}
    \mathrm{SDE} = \dfrac{\mathrm{SR}_\mathrm{peak} - \langle \mathrm{SR} \rangle}{\mathrm{sd} \left( \mathrm{SR} \right) } \ ,
\end{equation}
where $\mathrm{SR}_\mathrm{peak}$ is the highest value of the periodogram, $\langle \mathrm{SR} \rangle$ is the SR mean value, and $\mathrm{sd} \left( \mathrm{SR} \right)$ is the standard deviation of the SR values in the periodogram.

In order to narrow down the list of final candidates we applied the following cuts to the resulting parameters : \\ \\
\begin{tabular}{l c r c}
 -- $\mathcal{SNR}_\mathrm{T}$ &>& 7.5 &\\ 
 -- SparseBLS SDE &>& 6 &\\  
 -- Transit depth &<& 40 &mmag.   
\end{tabular} \\ \\
The last criterion was an attempt to avoid cases of eclipsing binaries, or Jovian exoplanets around M dwarfs,  which usually have depths larger than $40\,\mathrm{mmag}$. Those cases should be detectable by other tasks focusing on eclipsing binaries. We visually inspected the remaining $130$ candidates to look for clear transit-like features. We excluded $41$ candidates that did \textit{not} meet the following criteria:
\begin{itemize}
    \item Host star is a Main-Sequence star.
    \item Achromatic transit seen in $G_\mathrm{BP}$ and $G_\mathrm{RP}$.
    \item Transit is not V-shaped.
    \item No visible out-of-transit variability.
    \item No visible secondary eclipse. 
    \item No visible odd-even difference in \textit{TESS} photometry.
\end{itemize}

\subsection{Photometric validation}
In order to validate photometrically the remaining $89$ candidates, we searched for their light curves in the Full-Frame Image (FFI) photometry of \textit{TESS}. About half of them ($48$) were found more likely to be eclipsing binaries, or exhibited no transit in the \textit{TESS} data. Within the remaining $41$ candidates (to be published in \textit{Gaia} DR3, along with $173$ known exoplanets with visible transits in the photometry of \textit{Gaia}; Eyer et al., in prep.), we were able to find significant transit-like signals in the FFI data for $21$ stars. For these $21$ stars we used the FFI data to refine the transit parameters we had calculated during the preparation of the initial candidate set.

\section{Confirmed planets} \label{sec:confirmed}
We selected two of our leading candidates for confirmation by RV follow-up observations. Figs.~\ref{fig:30263_Phot} and \ref{fig:11079_Phot} show the normalized and combined photometry of \textit{Gaia} and \textit{TESS} for these two candidates, Gaia-1 and Gaia-2, along with the best-fitting models (Sect.~\ref{sec:analysis}), with $68\%$ confidence intervals. 
We detrended the \textit{Gaia} light curves using a simple linear fit. For the \textit{TESS} light curves we used the \texttt{Python} packages \texttt{Lightkurve} \citep{Lightkurve} and \texttt{tesscut} \citep{tesscut} to acquire the \textit{TESS} FFI photometry, and the \texttt{flatten()} method to detrend the raw data. 

The combined light curve of Gaia-1 contain $117$ measurements of \textit{Gaia} (in all three bands) and $952$ measurements of \textit{TESS}. We note the slight dilution in the \textit{TESS} light curve of Gaia-1. This may be due to the relatively large point spread function (PSF) of \textit{TESS}\footnote{In fact the relevant size in \textit{TESS} is the pixel response function -- PRF. See Section~6 of the \href{https://archive.stsci.edu/files/live/sites/mast/files/home/missions-and-data/active-missions/tess/_documents/TESS_Instrument_Handbook_v0.1.pdf}{TESS Instrument Handbook}.}, where the dilution is caused by the blended light of nearby stars included in the PSF. Gaia-2 is much brighter than its close neighbors and the blending effect is unnoticeable.

In the case of Gaia-2 we decided to exclude one measurement\footnote{$t=1841.3424606\, (\mathrm{BJD} - 2455197.5)$, marked with a green square in the top panel of Fig.~\ref{fig:11079_Phot}.}
from the \textit{Gaia} $G$ band that resided, after phase folding, in the middle of the transit, with a normalized flux of $f_\mathrm{norm} = 0.9965$. A closer look at this specific measurement reveals several indications for possible saturation. Furthermore, the $G_\mathrm{BP}$ and $G_\mathrm{RP}$ measurements, taken at almost the same time, do agree with the transit model.  Therefore and thanks to the validation by \textit{TESS} photometry, we were confident that this point was indeed an outlier, and that we could fit a more accurate model after excluding it. The combined light curve of Gaia-2 contains $394$ measurements of \textit{Gaia} (in all three bands) and $1183$ measurements of \textit{TESS}.

\subsection{Host stars} \label{sec:stellar}
The host stars Gaia-1 and Gaia-2 are listed in the TESS Input Catalog \citep[TIC;][]{TIC} with mass and radius estimates. We also used \textit{Gaia} data to estimate these values independently, using the \texttt{Python} package \texttt{isochrones} \citep{morton15}. This tool uses stellar evolution models, based on the distance and observable magnitudes in multiple bands. We used the \textit{Gaia} parallax and the three magnitudes ($G$, $G_\mathrm{BP}$, $G_\mathrm{RP}$), and obtained similar estimates, as listed in Table \ref{table:stellar}.
\begin{table*}[ht]
\centering
{\small \tabcolsep=20pt 
{\renewcommand{\arraystretch}{1.2}
\caption{Stellar parameter estimates for Gaia-1 and Gaia-2.}
\label{table:stellar}
 \begin{tabular}{ l l l l l} 
 Parameter & Units & \multicolumn{2}{c}{Value} & Source \\
 \hline
   & & \textbf{Gaia-1} & \textbf{Gaia-2} & \\
 \hline 
 Gaia SourceID & & 3026325426682637824 & 1107980654748582144 & Gaia EDR3 \\ 
 TIC ID & & 11755687 & 147797743 & TIC \\
 RA & deg &  $90.6436666838 \pm 3.4\cdot 10^{-9}$ &  $110.7353331726 \pm 2.2\cdot 10^{-9}$ & Gaia EDR3 \\
 DEC & deg & $-0.5771154808 \pm 3.0\cdot 10^{-9}$  & $67.2526599247 \pm 4.3\cdot 10^{-9}$ & Gaia EDR3 \\
 $G$ & mag &  $12.99192 \pm 0.00055$ & $11.20014 \pm 0.00043$ & Gaia \\
 $G_\mathrm{BP}$ & mag & $13.38862 \pm 0.00051$ & $11.54127 \pm 0.00026$ & Gaia \\
 $G_\mathrm{RP}$ & mag & $12.41065 \pm 0.00051$ & $10.69055 \pm 0.00028$ & Gaia \\
 $V$ & mag & $13.242 \pm 0.092$ & $11.277 \pm 0.008$ & TIC \\
 Parallax & mas & $2.715 \pm 0.015$ & $4.826 \pm 0.023 $ & Gaia EDR3 \\
 \\
 $T_{\mathrm{eff}}$ & K & $5470\pm110$ & $5720\pm84$ & \texttt{isochrones} \\
 $M_*$&$M_\odot$ & $0.949\pm0.066$ & $1.000\pm0.095$ & \texttt{isochrones} \\
 $R_*$&$R_\odot$ & $0.952\pm0.025$ & $1.064\pm0.031$ & \texttt{isochrones} \\
 $\rho_*$ & $\textrm{kg\,m}^{-3}$ & $1558 \pm 170$ & $1170 \pm 160$ & \texttt{isochrones} \\
 \\
 $T_{\mathrm{eff}}$ & K & $5370 \pm 140$ & $5720 \pm 130$ & TIC \\
 $M_*$&$M_\odot$ & $0.93 \pm 0.12$  & $1.02 \pm 0.13$ & TIC \\
 $R_*$&$R_\odot$ & $0.962 \pm 0.054$  & $1.088 \pm 0.053$ & TIC \\
 $\rho_*$ & $\textrm{kg\,m}^{-3}$ & $1480 \pm 310$ & $1120 \pm 220$ & TIC \\
 
 \end{tabular}}}
\end{table*}

\begin{figure}
	\includegraphics[width=1\linewidth]{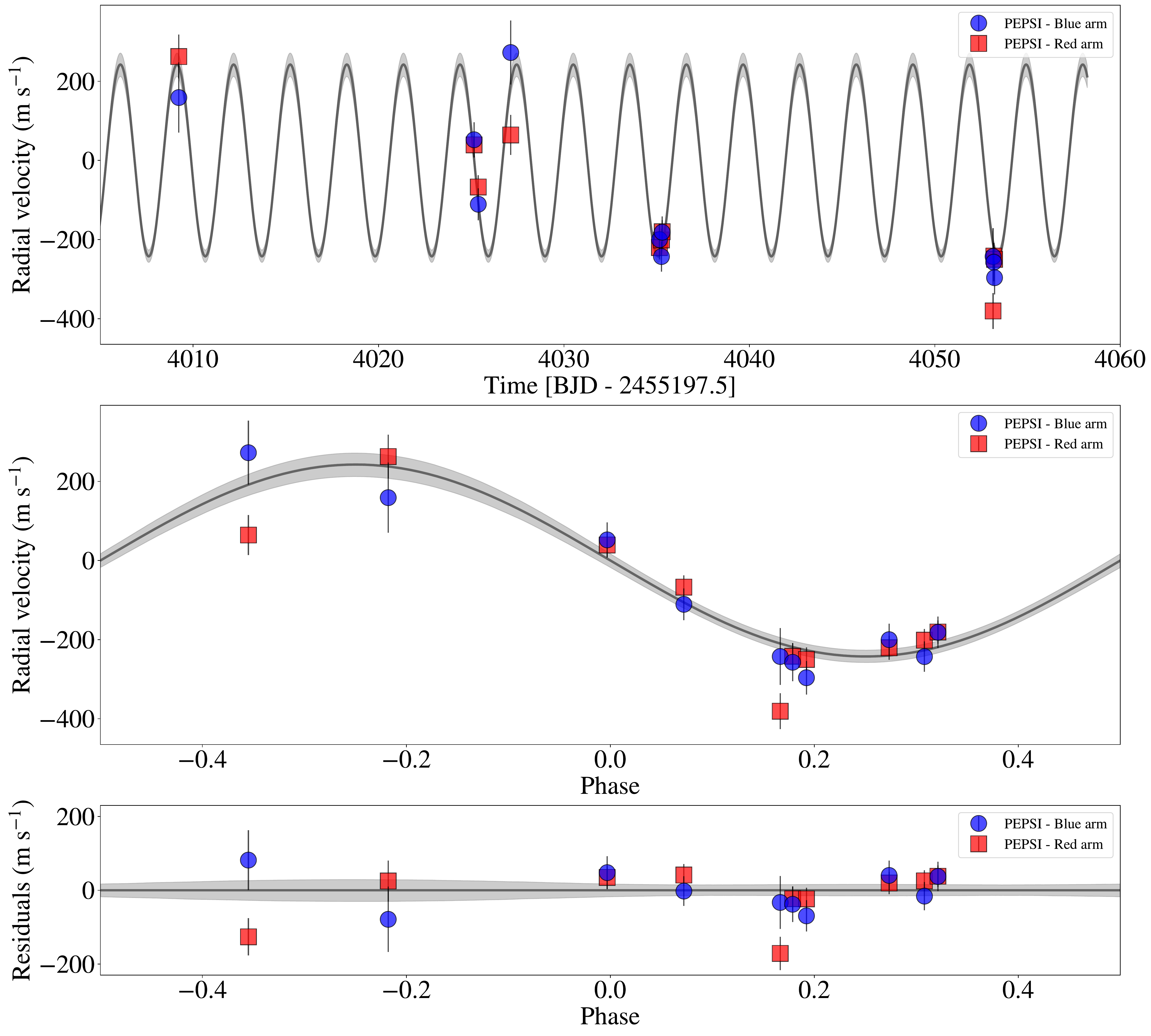}
	\caption{(\textit{Top}) PEPSI RV measurements of Gaia-1, on top of the best-fitting RV curve (solid line) derived by \texttt{juliet}. We subtracted the systemic velocity, listed in Table~\ref{table:posteriors}. (\textit{Middle}) Phase-folded RV curves according to the period listed in Table~\ref{table:posteriors}. (\textit{Bottom}) Residuals with no visible systematic variation.}
	\label{fig:30263_RV} 
\end{figure}
\begin{figure}
	\includegraphics[width=1\linewidth]{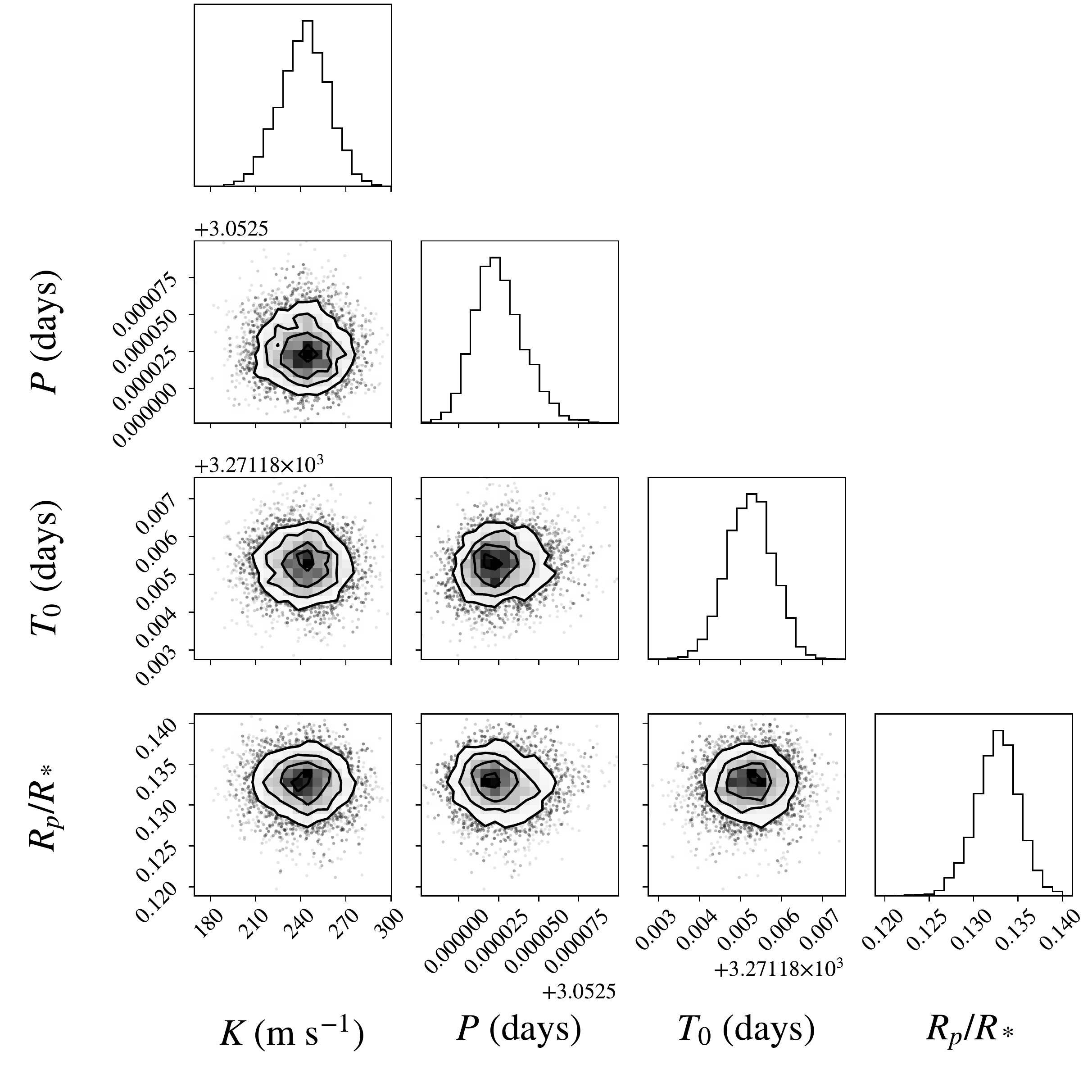}
	\caption{Gaia-1b: Corner plot of the posterior distributions of the main system parameters, assuming circular orbit.}
	\label{fig:30263_Corner} 
\end{figure}

\subsection{Radial velocities} \label{sec:RV}
We obtained high-resolution spectra for the two candidates, using the spectrograph PEPSI \citep{2015PEPSI}, on the Large Binocular Telescope (LBT). 
The PEPSI spectrograph has two arms, blue and red, with six cross-dispersers for a full optical coverage in the range of $383-907$\,nm. In this work we used the high-resolution configuration with $R=50\,000$ using cross-dispersers 2 (blue arm) and 4 (red arm), covering the wavelength ranges $426-480$ and $544-627$\,nm, respectively \citep{2015PEPSI}. 
The PEPSI pipeline produces a one-dimensional spectrum for each order, wavelength-calibrated using a ThAr lamp, continuum-normalized and corrected for solar barycentric motion. The cross-correlation function (CCF) was calculated independently for each spectral order. The radial velocities and their uncertainties were derived from the combined CCF, according to \cite{2003MNRAS.342.1291Z}, using the \texttt{Python} package \texttt{SPARTA} \citep{SPARTA}. 
The RV modulations are detailed in Table~\ref{table:RVs}, with the systemic velocity subtracted (See Table~\ref{table:posteriors}).
\begin{table}
\centering
\caption{\\PEPSI RV measurements of Gaia-1 and Gaia-2, extracted with \texttt{SPARTA}. Systemic velocities of the circular orbits were removed (See Table \ref{table:posteriors}).}
\label{table:RVs}
{\small
{\renewcommand{\arraystretch}{1.1}
 \begin{tabular}{ l r r } 
 \textbf{Gaia-1} \\
 \hline 
 Time ($\mathrm{BJD} - 2455197.5$) & $\mathrm{RV_{Blue}}\,(\mathrm{m\,s^{-1}})$ & $\mathrm{RV_{Red}}\,(\mathrm{m\,s^{-1}})$ \\
 \hline 
 $4009.231299$ & $159 \pm 89$ & $262 \pm 55$ \\
$4025.149340$ & $52 \pm 44$   & $39 \pm 32$ \\
$4025.378759$ & $-111 \pm 40$ & $-67 \pm 30$ \\
$4027.127925$ & $273 \pm 81$  & $64 \pm 51$ \\
$4035.150587$ & $-200 \pm 40$ & $-220 \pm 30$ \\
$4035.256314$ & $-243 \pm 38$ & $-201 \pm 29$ \\
$4035.296691$ & $-181 \pm 40$ & $-180 \pm 29$ \\
$4053.139787$ & $-243 \pm 72$ & $-380 \pm 45$ \\
$4053.176891$ & $-257 \pm 48$ & $-242 \pm 33$ \\
$4053.218314$ & $-296 \pm 43$ & $-250 \pm 30$ \\
 \\
 \textbf{Gaia-2} \\
 \hline 
 Time ($\mathrm{BJD} - 2455197.5$) & $\mathrm{RV_{Blue}}\,(\mathrm{m\,s^{-1}})$ & $\mathrm{RV_{Red}}\,(\mathrm{m\,s^{-1}})$ \\
 \hline 
 $4009.242638$ & $-111 \pm 30$  & $-89 \pm 26$ \\
$4025.160356$  & $55 \pm 26$    & $50 \pm 14$ \\
$4025.405016$  & $90 \pm 25$    & $86 \pm 14$ \\
$4035.167108$  & $-67 \pm 25$   & $-48 \pm 14$ \\
$4035.278306$  & $-2 \pm 25$    & $-21 \pm 14$ \\
$4053.156642$  & $-144 \pm 27$  & $-164 \pm 14$ \\
$4053.197203$  & $-147 \pm 25$  & $-141 \pm 14$ \\
$4053.238726$  & $-140 \pm 25$  & $-128 \pm 14$ \\
 \end{tabular}}}
\end{table}

\subsection{Analysis} \label{sec:analysis}
We performed joint analyses incorporating the photometry of \textit{Gaia} and \textit{TESS}, together with the RV data of the red and blue arms of PEPSI as effectively two different instruments, using the \texttt{Python} package \texttt{juliet} \citep{JULIET}. \texttt{juliet} uses \texttt{batman}, \citep{batman} for modeling the transit light curve and \texttt{radvel} \citep{radvel} for modeling the RV curve. \texttt{juliet} uses several parametrization schemes that allow the sampling of the parameter space while maintaining the physical validity of the model. \citet{Espinoza_Sampling} and \citet{10.1093/mnras/stt1435} provide additional details of the sampling schemes and the parametrization. We used the Dynamic Nested Sampling method \citep[\texttt{dynesty};][]{dynesty} to get parameter posterior  estimates, along with the Bayesian log evidence ($\ln Z$) for each model, useful for comparing different models. According to \cite{Trotta2008}, a difference of $\Delta \ln Z < 1$ means the two models should be considered statistically indistinguishable, while $\Delta \ln Z > 5$ suggests strong evidence in favor of the model with the larger value of $\ln Z$.
Besides selecting \texttt{dynesty} for the sampling method, all other parameters of the \texttt{fit()} method of \texttt{juliet} were left in their default values.
Convergence is achieved when the program fails to improve its $\ln Z$ value by $0.5$ in one complete iteration.

Given the relatively low number of RV measurements, we decided to assume a circular orbit and fix the eccentricity at zero, as expected for planets with such short periods \citep[e.g.][]{2003ASPC..294..213W}.
We used similar priors for the various parameters as those used by \cite{2020MNRAS.491.2982E}, which we detail in Table~\ref{table:priors}. We have set all jitter terms $\sigma^{}_{\omega}$ to zero, as well as the flux offset terms $M$, since we used normalized light curves in this analysis. The mean values for the priors of the period and time of mid-transit were estimated based on the results of the preliminary analysis of the photometry.

\begin{figure}
	\includegraphics[width=1\linewidth]{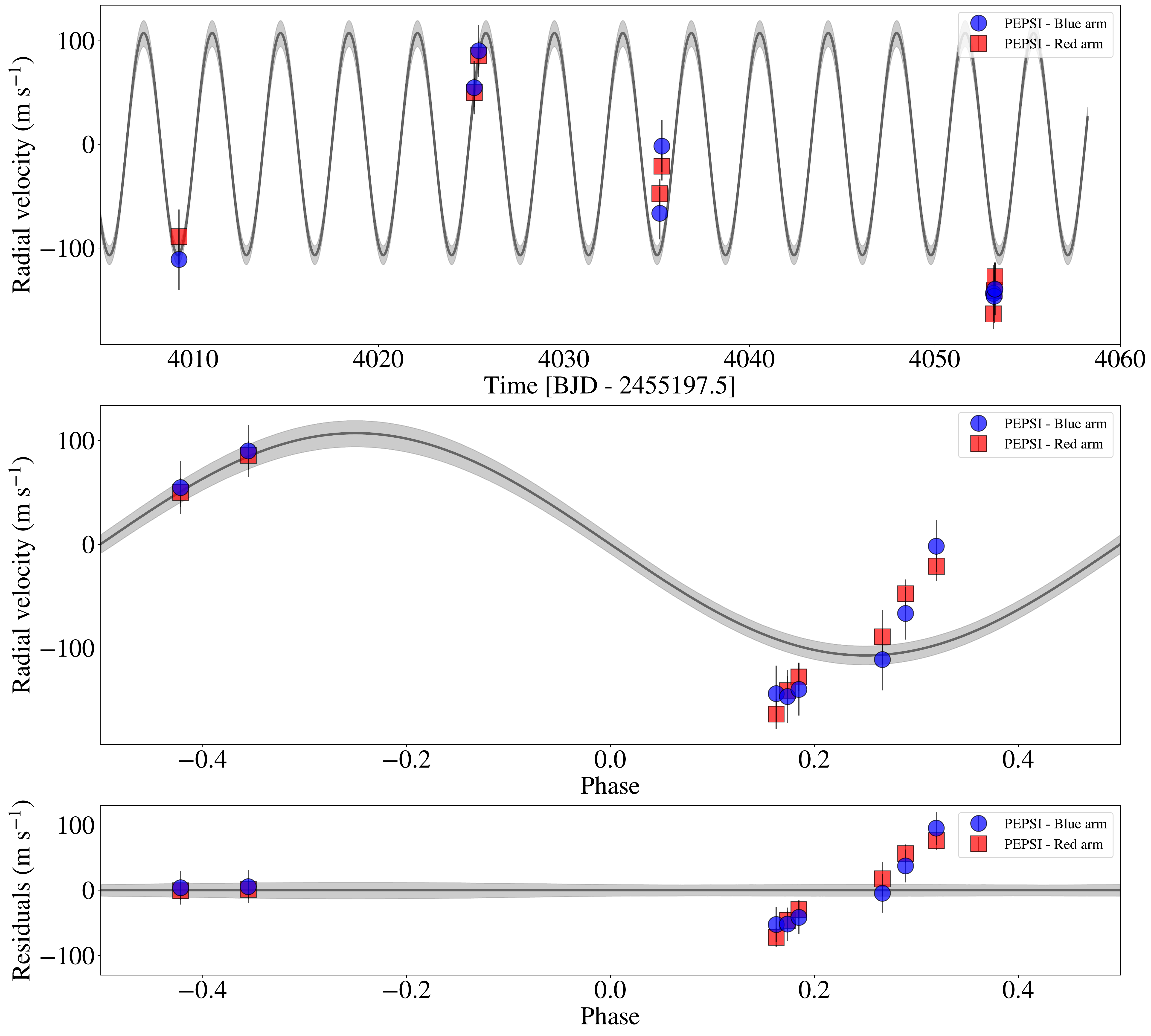}
	\caption{(\textit{Top}) PEPSI RV measurements of Gaia-2, on top of the best-fitting RV curve (solid line) derived by \texttt{juliet}. We subtracted the systemic velocity, listed in Table~\ref{table:posteriors}. (\textit{Middle}) Phase-folded RV curves according to the period listed in Table~\ref{table:posteriors}. (\textit{Bottom}) Residuals showing a possible systematic variation, perhaps due to some eccentricity of the orbit.}
	\label{fig:11079_RV} 
\end{figure}
\begin{figure}
	\includegraphics[width=0.95\linewidth]{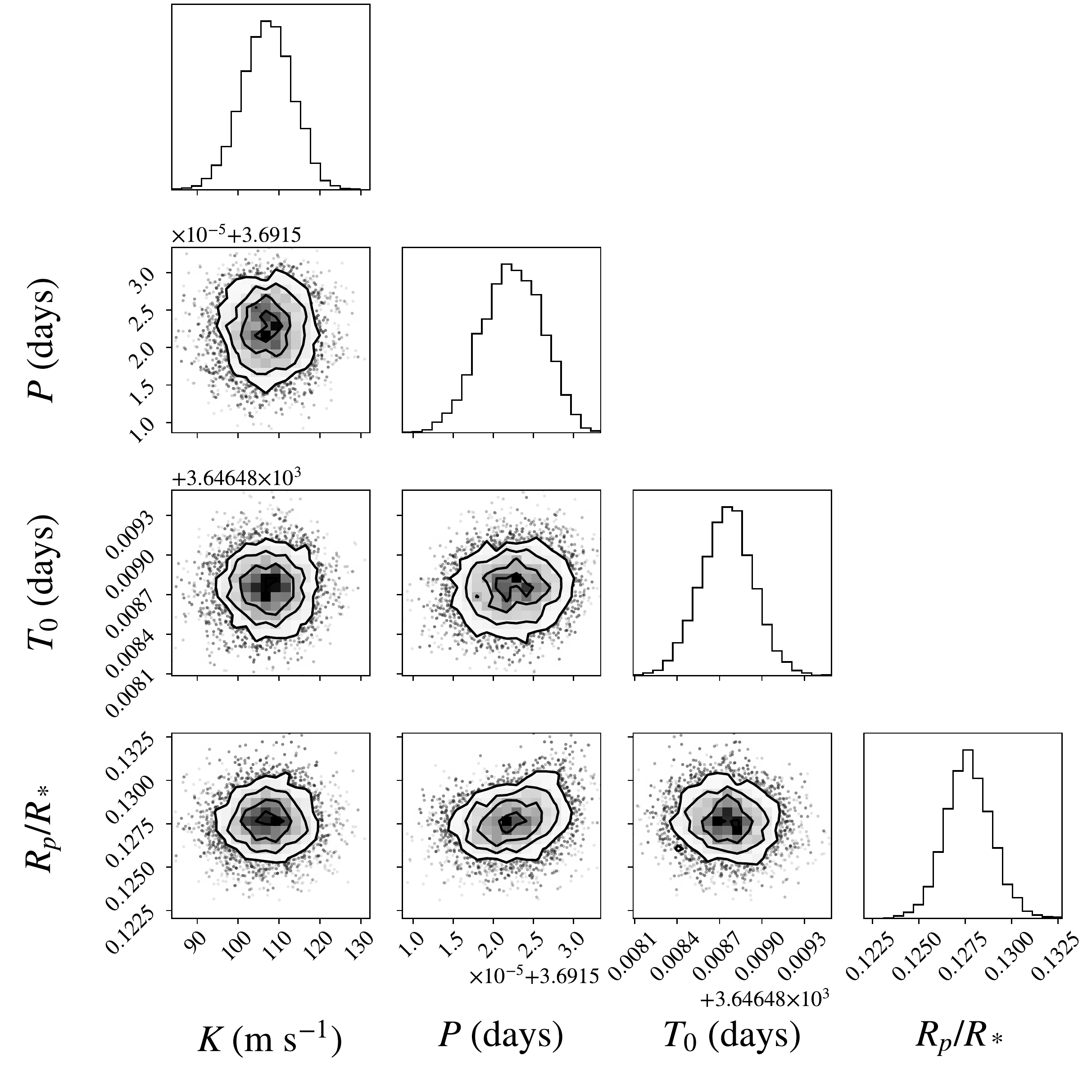}
	\caption{Gaia-2b: Corner plot of the posterior distribution of the main system parameters, assuming circular orbit.}
	\label{fig:11079_Corner} 
\end{figure}

\begin{table*}[ht]
{\small \tabcolsep=10pt 
\caption{Prior distributions for the joint photometry and RV analysis of Gaia-1b and Gaia-2b. We denote uniform distributions between $a$ and $b$ as $\mathcal{U}(a,b)$ and normal distributions with mean $\mu$ and variance $\sigma^2$ as $\mathcal{N}(\mu , \sigma^2) $.}
\label{table:priors}
{\renewcommand{\arraystretch}{1.2}
 \begin{tabular}{ l l l l l } 
 \hline 
 Parameter & Description & Units & \multicolumn{2}{c}{Prior} \\
 \hline 
  & & & \textbf{Gaia-1b} & \textbf{Gaia-2b} \\
 \hline 
 $P$ & Period & days & $\mathcal{N}(3.052503 , 0.01^2)$ & $\mathcal{N}(3.691508 , 0.01^2)$ \\ 
 $T_0$ & Time of mid-transit & BJD - 2455197.5  & $\mathcal{N}(3271.23705 , 0.1^2)$ & $\mathcal{N}(3646.43546, 0.1^2)$ \\ 
 $K$ & Semi-amplitude of the radial velocity & $\textrm{m\, s}^{-1}$ & $\mathcal{U}(0 , 500)$ & $\mathcal{U}(0 , 500)$ \\ 
 $e$ & Eccentricity & - & 0 - fixed & 0 - fixed ; $\mathcal{U}(0 , 0.95)^i$\\ 
 $\omega$ & Argument of periastron & degrees & 0 - fixed & 0 - fixed ; $\mathcal{U}(0 , 360)^i$ \\ 
 $\rho_*$ & Stellar mass mean density & $\textrm{kg\,m}^{-3}$  & $\mathcal{N}(1558 , 170^2)$ & $\mathcal{N}(1173 , 164^2)$\\\\
 $r_1,r_2$ & Parametrization of $p, b \, ^{ii}$ & - & $\mathcal{U}(0 , 1)$ & $\mathcal{U}(0 , 1)$\\
 $q_{1,\mathrm{Gaia}},q_{2,\mathrm{Gaia}}$ & Limb-darkening parametrization$^{iii}$ for \textit{Gaia} & - & $\mathcal{U}(0 , 1)$ & $\mathcal{U}(0 , 1)$\\
 $D_{\mathrm{Gaia}}$ & Dilution factor for \textit{Gaia} & - & 1 - fixed & 1 - fixed \\ 
 \\
 $q_{1,\mathrm{TESS}},q_{2,\mathrm{TESS}}$ & Limb-darkening parametrization$^{iii}$ for \textit{TESS} & - & $\mathcal{U}(0 , 1)$ & $\mathcal{U}(0 , 1)$\\
 $D_{\mathrm{TESS}}$ & Dilution factor for \textit{TESS} & - & $\mathcal{U}(0.1 , 1.0)$ & $\mathcal{U}(0.1 , 1.0)$ \\ 
 \\
 $\gamma$ & Relative center-of-mass velocity for PEPSI$^{iv}$ & $\textrm{m\, s}^{-1}$ & $\mathcal{U}(-500 , 500)$ & $\mathcal{U}(-500 , 500)$ \\ 
 \\
 \end{tabular}}
 
 $^{i}$ Separate analysis, allowing eccentricity.\\ 
 $^{ii}$ Described by \cite{Espinoza_Sampling}, $p = {R_p}/{R_*}$ is the planetary to stellar radius ratio, and $b = \left ( a/R_* \right ) \cos i\,$ is the impact parameter.\\
 $^{iii}$ Described by \cite{10.1093/mnras/stt1435}.\\
 $^{iv}$ Around a middle value of $-37\,750$ for Gaia-1b and $-36\,000$ for Gaia-2b.
 }
\end{table*}

The posterior medians and $68\%$ confidence intervals of the system parameters resulting from the \texttt{juliet} analyses are detailed in Table~\ref{table:posteriors}, accompanied by corner plots  \citep{corner} for the main parameters in Figs.~\ref{fig:30263_Corner} and \ref{fig:11079_Corner}. We also used $68\%$ confidence intervals for the RV models in Figs.~\ref{fig:30263_RV} and \ref{fig:11079_RV}.
All two-dimensional histograms in the corner plots have four contour lines representing levels of $(0.5,1.0,1.5,2.0)$ sigmas.

\subsection{Gaia-1b}
The RVs of Gaia-1 seem to closely trace a sine curve (Fig.~\ref{fig:30263_RV}), as expected for a circular Keplerian orbit. Based on the estimated stellar parameters, we estimate the mass and radius of the transiting object to be 
$M_\mathrm{p} = 1.68 \pm 0.11 \, M_{\mathrm{J}}$, 
$R_\mathrm{p} = 1.229 \pm 0.021 \, R_{\mathrm{J}}$, consistent with a possibly inflated hot Jupiter. 
For completeness, we tried to fit an eccentric orbit, but found no statistical evidence supporting an eccentric model ($\Delta \ln Z < 1$).
When comparing to a model with no planet we got a value of $\Delta \ln Z = 109$, suggesting strong evidence for the existence of Gaia-1b. 
The residuals in Fig.~\ref{fig:30263_Phot} show a possible systematic variation, suggesting some out-of-transit variability, possibly due to a more massive companion. No such variability was observed in the photometry of \textit{TESS}. Furthermore, the scatter seem not to be consistent among the three bandpasses of \textit{Gaia} photometry, and the joint RV and photometry analysis suggests a planetary companion. We therefore concluded there was no substantial evidence for this variability.

\subsection{Gaia-2b}
\begin{figure}
	\includegraphics[width=1\linewidth]{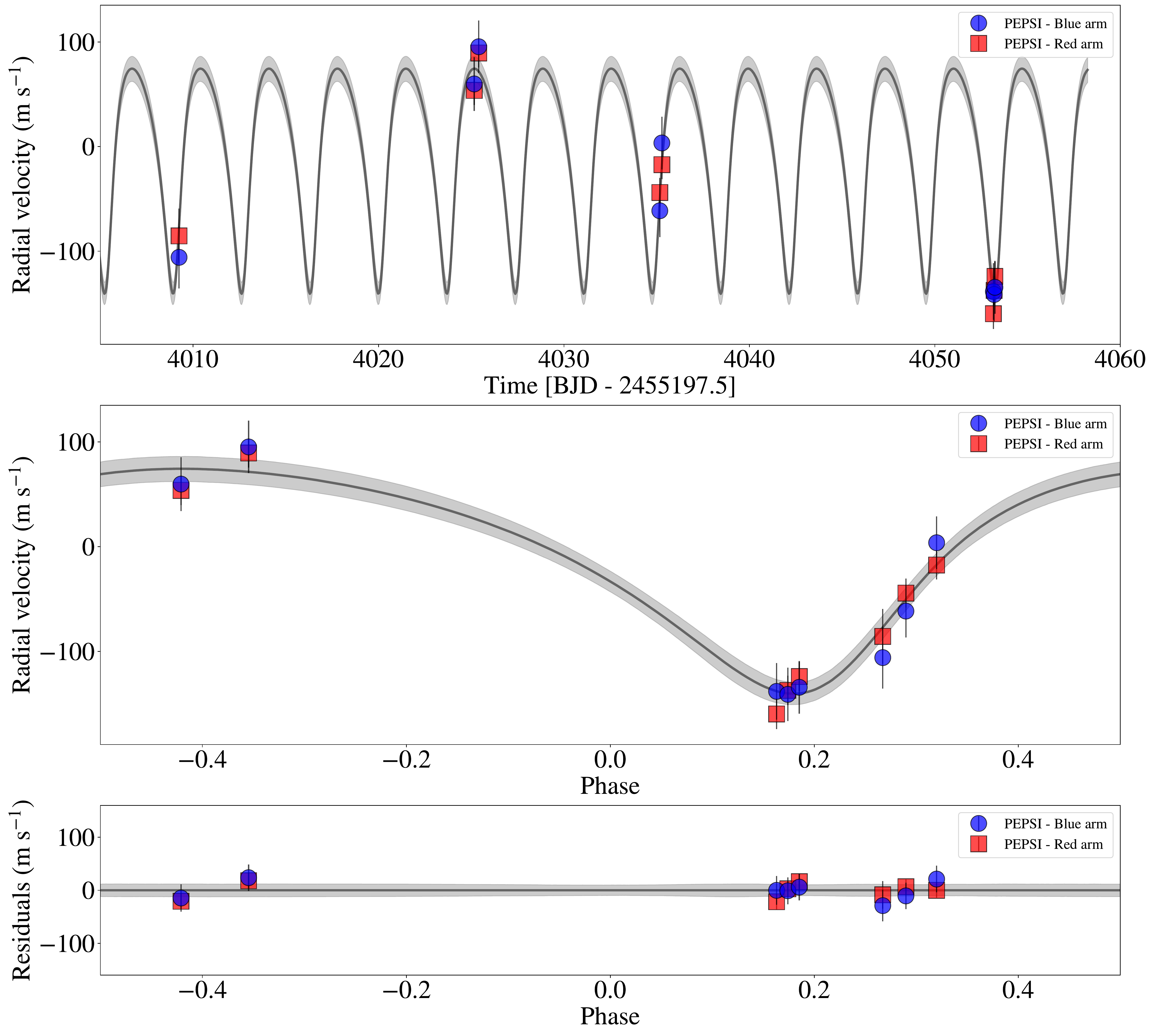}
	\caption{(\textit{Top}) RV measurements of Gaia-2, on top of the best-fitting eccentric model (solid line). We subtracted the systemic velocity, listed in Table~\ref{table:posteriors}. (\textit{Middle}) Phase-folded RV curves according to the period listed in Table~\ref{table:posteriors}. (\textit{Bottom}) Residuals with no visible systematic variation.}
	\label{fig:11079_RV_ecc} 
\end{figure}

\begin{figure}
	\includegraphics[width=1\linewidth]{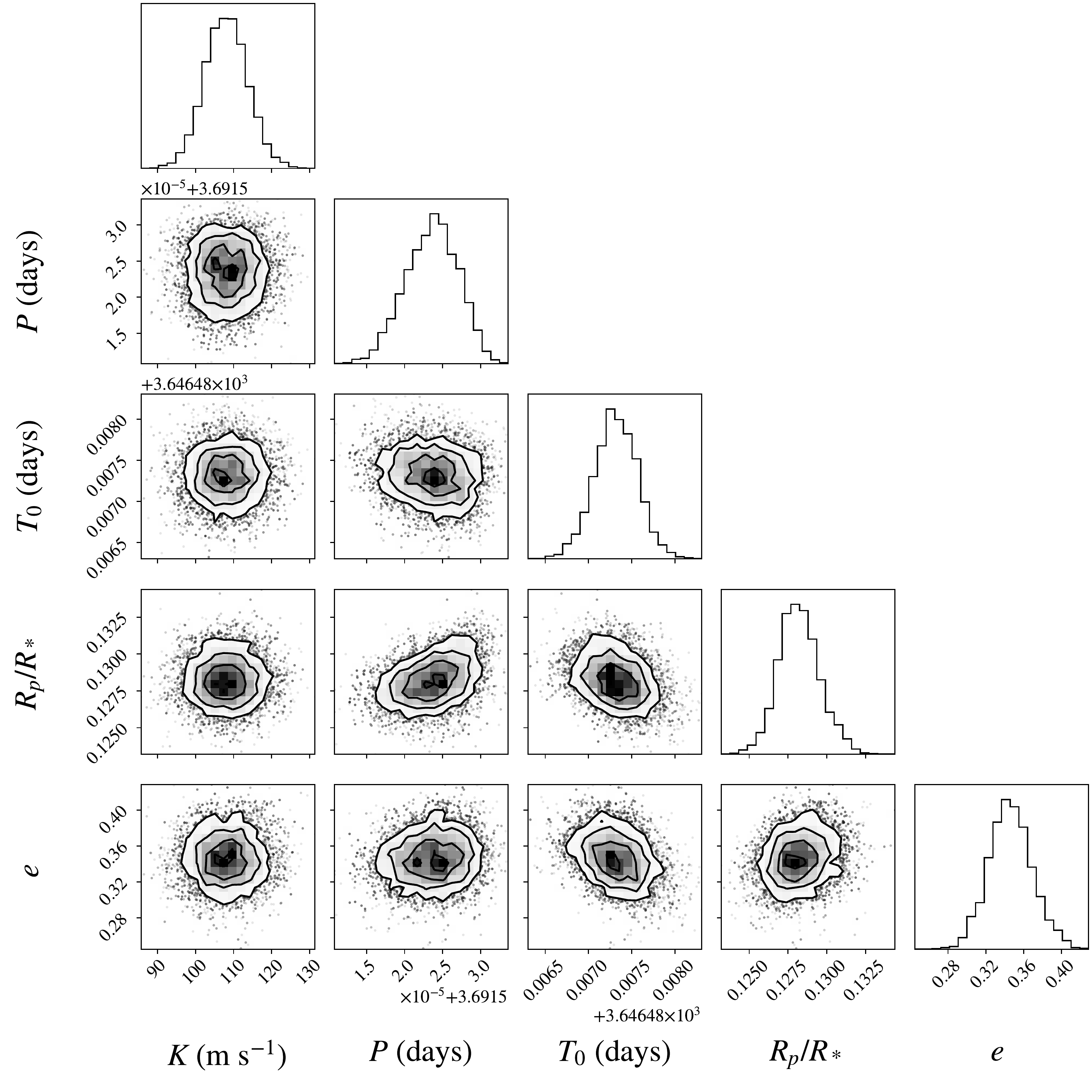}
	\caption{Gaia-2b: Corner plot of the posterior distribution of the main system parameters for a non-circular orbit.}
	\label{fig:11079_Corner_ecc} 
\end{figure}

The phase coverage of the Gaia-2 RV measurements is suboptimal (Fig.~\ref{fig:11079_RV}), especially around phase zero, and the phase-folded RV curve seems to suggest a potentially eccentric orbit. We therefore performed an additional analysis allowing non-zero eccentricity, with a uniform prior distribution $\mathcal{U}(0,0.95)$, resulting in an estimate for the eccentricity of $e= 0.346 \pm 0.023$ with $\Delta \ln Z = 48$ over the circular orbit. The RVs with the eccentric model are shown in Fig.~\ref{fig:11079_RV_ecc}, and the posterior distributions and estimates for the main orbital parameters are given in Fig.~\ref{fig:11079_Corner_ecc} and Table~\ref{table:posteriors}. Despite the strong statistical evidence, such an eccentric orbit would be very surprising given the proximity of the planet to its host star. We therefore attempted also to fit a circular orbit with a constant slope in the RV curve, using a uniform prior distribution $\mathcal{U}(-300,300) \,\mathrm{m\,s^{-1}\,d^{-1}}$, resulting in an estimate for the RV slope of $RV_{slope}= -2.9 \pm 0.39 \,\mathrm{m\,s^{-1}\,d^{-1}}$ with $\Delta \ln Z = 18$ over the no-slope, circular model. Statistically, the eccentric orbit seems to be preferable, but given the small number of measurements and their uncertainties, we decided to keep the more plausible circular orbit, and wait for future RV measurements to better constrain this system.

We then estimate the mass and radius of the transiting object to be $M_\mathrm{p} = 0.817 \pm 0.047 \, M_{\mathrm{J}}$, 
$R_\mathrm{p} = 1.322 \pm 0.013 \, R_{\mathrm{J}}$, 
also consistent with a potentially inflated hot Jupiter. When comparing to a model with no planet we got a value of $\Delta \ln Z = 133$, suggesting strong evidence for the existence of Gaia-2b.
\begin{table*}
{\tabcolsep=10pt 
{\renewcommand{\arraystretch}{1.2}
\caption{Posterior estimates for Gaia-1b and Gaia-2b.}
\label{table:posteriors}
 \begin{tabular}{ l l l l l} 
 \hline
 Parameter & Units & \multicolumn{3}{c}{Value}\\
 \hline 
 & & \textbf{Gaia-1b} & \multicolumn{2}{c}{\textbf{Gaia-2b}} \\
 \hline
 & & & \textbf{Circular Orbit} & \textbf{Eccentric Orbit} \\
 $P          $   & days                                      &    $3.052524 \pm 1.7\cdot 10^{-5}$          &  $3.6915224 \pm 3.9\cdot 10^{-6}$   &  $3.6915237 \pm 3.5\cdot 10^{-6}$     \\
 $T_0         $  & BJD - 2455197.5                           &    $3271.18524 \pm 5.7\cdot 10^{-4}$        &  $3646.48875 \pm 1.9\cdot 10^{-4}$  &  $3646.48731 \pm 2.5\cdot 10^{-4}$    \\
 $r_1         $  & -                                         &    $0.773 ^{+0.028} _{-0.037}$  		       &  $0.806 \pm 0.015$  		         &  $0.816 \pm 0.015$                  \\
 $r_2         $  & -                                         &    $0.124 \pm 0.0023$  		               &  $0.1277 \pm 0.0013$  		         &  $0.1282 \pm 0.0013$                   \\
 $q_{1,\mathrm{Gaia}}  $  & -                                &    $0.24 ^{+0.25} _{-0.14}$ 		           &  $0.125 ^{+0.125} _{-0.078}$  	   	 &  $0.138 ^{+0.123} _{-0.080}$           \\
 $q_{2,\mathrm{Gaia}}  $  & -                                &    $0.42 ^{+0.34} _{-0.26}$ 		           &  $0.39 ^{+0.33} _{-0.26}$  	   	 &  $0.43 ^{+0.31} _{-0.26}$             \\
 $q_{1,\mathrm{TESS}}  $  & -                                &    $0.64 ^{+0.23} _{-0.30}$  		           &  $0.67 ^{+0.19} _{-0.16}$  	     &  $0.64 \pm 0.18$                      \\
 $q_{2,\mathrm{TESS}}  $  & -                                &    $0.36 ^{+0.29} _{-0.22}$                 &  $0.24 ^{+0.16} _{-0.13}$  	     &  $0.20 ^{+0.16} _{-0.12}$               \\
 $D_{\mathrm{TESS}}    $  & -                                &    $0.767 \pm 0.036$  		               &  $0.966 ^{+0.019} _{-0.022}$  		 &  $0.969 ^{+0.018} _{-0.021}$            \\
 $\rho_*      $  & $\textrm{kg\,m}^{-3}$                     &    $1460 ^{+160} _{-140}$	           &  $848 ^{+60} _{-53}$          &  $1080 ^{+140} _{-110}$         \\
 $\gamma_{Blue} $  & $\textrm{m\, s}^{-1}$                   &    $-37675 \pm 18$                          &  $-35994 \pm 9$                   &  $-35999 \pm 10$                      \\
 $\gamma_{Red} $  & $\textrm{m\, s}^{-1}$                    &    $-37915 \pm 14$                          &  $-35967 \pm 6$                   &  $-35971 \pm 6$                       \\
 $K          $  & $\textrm{m\, s}^{-1}$                      &    $243 \pm 16$  		                   &  $107.1 \pm 6.2$  		             &  $108.0 \pm 5.6$                        \\
 Inclination   & degrees                                     &    $85.73 ^{+0.47} _{-0.41}$                &  $85.21 \pm 0.25$                   &  $85.66 \pm 0.31$                     \\
 $a$ (semi-major axis) & AU                                  &    $0.04047 \pm 9.4\cdot 10^{-4}$           &  $0.0467 \pm 0.0015$                &  $0.0467 \pm 0.0015$                  \\
 $e$  & -                                                    &    -                                        &  -                                  &  $0.346 \pm 0.023$                    \\
 $\omega$  & degrees                                         &    -                                        &  -                                  &  $206.2 ^{+6.2} _{-6.7}$              \\
 $M_\mathrm{p}$      & $M_{\textrm{J}}$                      &    $1.68 \pm 0.11$  		                   &  $0.817 \pm 0.047$  		         &  $0.773 \pm 0.041$             \\
 $R_\mathrm{p}$      & $R_{\textrm{J}}$                      &    $1.229 \pm 0.021$  		               &  $1.322 \pm 0.013$  		         &  $1.327 \pm 0.014$                    \\
 $\ln Z$  & -                                                &    $5518.7$                                 &  $8237.3$                           &  $8285.6$                             \\
 $\ln Z$ without a planet & -                                 &    $5409.6$                                 &  $8104.3$                           &  -                                    
 \end{tabular}}}                 
\end{table*}

\section{Conclusions} \label{sec:conc}
In this paper we described the method used by DPAC to find the first batch of transiting exoplanet candidates based on \textit{Gaia} photometry. For the first batch of candidates we aimed at detecting the easiest cases, i.e.\ hot Jupiters -- giant planets that orbit their stars in short periods of a few days at most. Transiting hot Jupiters are relatively easy to detect because they exhibit relatively deep transits and the transits duty cycles are large. For the reasons mentioned in Sect.~\ref{sec:methods}, the presented search does not pretend to be exhaustive, and is far from exploiting the full detection capability of \textit{Gaia}. We therefore do not attempt to estimate the completeness of this search, since its statistical value is only limited at this early stage.

However, even at this point it is clear that the circumstances in the future data releases (DR4 and DR5) will allow more detections of transiting exoplanets. First, the training set used to train the classifier will be based on more data, leading to a better selection of initial candidates. More importantly, since the number of measurements in each light curve will be larger, \textit{Gaia} photometry is bound to capture more transits, thus enhancing significantly the ability of the BLS approach to identifying them. Given the longer observation time baseline and the larger number of observations, one can predict that DR4 and DR5 will include larger sets of candidates, possibly also covering a wider range of orbital periods.

The confirmation of the two planets Gaia-1b and Gaia-2b serves to validate the presented search methodology. For Gaia-2b we could not rule out a more eccentric orbit, or an additional massive object that induces an RV slope, due to insufficient phase coverage, and it will probably be resolved by future RV measurements. Even without a detectable RV slope, an eccentric orbit can potentially still be the result of the presence of a third massive object in the system, which induces a non-zero eccentricity of the planet \citep{MazSha1979}. In any case, even when allowing for eccentricity or an RV slope in our fits, the estimated mass of the transiting object was always less than $1.5\,M_{\textrm{J}}$, well within the planetary regime.

The capability of \textit{Gaia} to photometrically detect transiting exoplanets has often been questioned. Nevertheless, recognizing the potential, several authors have tried to estimate \textit{Gaia} yield of transiting exoplanets \citep{HOG2002,Robichon2002_2,Dzigan_2012}, based on assumptions concerning Galactic models, planet frequency, and \textit{Gaia} photometric performance. The \textit{Gaia} mission has been given an indicative approval for an extension until the end of 2025
\footnote{\url{https://sci.esa.int/web/director-desk/-/extended-operations-confirmed-for-science-missions}}, probably increasing significantly the detection potential. \textit{TESS} is performing its own all-sky survey for transiting exoplanets, but its mode of operation is focusing on short-period transits. \textit{Gaia} is monitoring a larger sample of stars than \textit{TESS}, and with the longer observing time span it potentially can detect long-period planets. Thus, having established that it can detect planetary transits, \textit{Gaia} will complement the capabilities of \textit{TESS}.

\begin{acknowledgements}
We would like to thank the anonymous referee for the insightful and helpful comments that greatly improved the analyses done in this project. This work has made use of data from the European Space Agency (ESA) mission Gaia (\url{https://www.cosmos.esa.int/gaia}), processed by the Gaia Data Processing and Analysis Consortium (DPAC, \url{https://www.cosmos.esa.int/web/gaia/dpac/consortium}). 
Funding for the DPAC has been provided by national institutions, some of which participate in the \textit{Gaia} Multilateral Agreement,
which include, for Switzerland, the Swiss State Secretariat for Education, Research and Innovation through the Activit\'{e}s Nationales Compl\'{e}mentaires (ANC).
This work was supported by a grant from the Tel Aviv University Center for AI and Data Science (TAD) and
by the Ministry of Science, Technology and Space, Israel (Grant 3-18143).
This research has made use of the NASA Exoplanet Archive, which is operated by the California Institute of Technology, under contract with the National Aeronautics and Space Administration under the Exoplanet Exploration Program. 
We would like to thank Andrea Rossi (INAF-OAS Bologna) for carrying out the LBT observations and operating the PEPSI spectrograph. The LBT is an international collaboration among institutions in the United States, Italy and Germany. LBT Corporation partners are: The University of Arizona on behalf of the Arizona Board of Regents; Istituto Nazionale di Astrofisica, Italy; LBT Beteiligungsgesellschaft, Germany, representing the Max-Planck Society, The Leibniz Institute for Astrophysics Potsdam, and Heidelberg University; The Ohio State University, representing OSU, University of Notre Dame, University of Minnesota and University of Virginia.
\texttt{Python} libraries used: \texttt{Matplotlib} \citep{matplotlib}, \texttt{NumPy} \citep{numpy}, \texttt{AstroPy} \citep{astropy}, \texttt{Lightkurve} \citep{Lightkurve}, \texttt{juliet} \citep{JULIET}, \texttt{batman} \citep{batman}, \texttt{radvel} \citep{radvel}, \texttt{dynesty} \citep{dynesty}, \texttt{SPARTA} \citep{SPARTA} and \texttt{Pandas} \citep{Pandas}.
\end{acknowledgements}
\bibliographystyle{aa}
\bibliography{Main.bib}
\end{document}